\documentstyle[11pt,aaspp4,psfig]{article}
\newcommand{\simgt}{\lower.5ex\hbox{$\; \buildrel > \over \sim \;$}}
\newcommand{\simlt}{\lower.5ex\hbox{$\; \buildrel < \over \sim \;$}}
\newcommand{\gequal}{\lower.5ex\hbox{\tiny$\stackrel{\textstyle >}{=}$}}
\newcommand{\himpc}{{\hbox {$h^{-1}$}{\rm Mpc}} }
\def\himsun{{h^{-1}M_\odot}}
\def\msun{{M_\odot}}
\def\kms{\,{\rm {km\, s^{-1}}}}
\begin{document}

\title{Confronting cold dark matter cosmologies \\
with strong clustering of Lyman break galaxies at $z\sim3$}

\bigskip

\author{Y.P. Jing and Yasushi Suto}

\bigskip
\bigskip

\affil{Department of Physics and
Research Center for the Early Universe (RESCEU)\\
School of Science, The University of Tokyo, Tokyo 113, Japan.\\
~\\
e-mail: jing@utaphp2.phys.s.u-tokyo.ac.jp, ~
suto@phys.s.u-tokyo.ac.jp}

\received{1997 October 9} \accepted{1997 November}

\begin{abstract}
  We perform a detailed analysis of the statistical significance of a
  concentration of Lyman break galaxies at $z \sim 3$ recently
  discovered by Steidel et al., using a series of N-body
  simulations with $N=256^3$ particles in a $(100\himpc)^3$ comoving
  box.  While the observed number density of Lyman break galaxies at
  $z\sim3$ implies that they correspond to systems with dark matter
  halos of $\simlt 10^{12}M_\odot$, the resulting clustering of such
  objects on average is not strong enough to be reconciled with the
  concentration if it is fairly common; we predict one similar
  concentration approximately per ($6\sim 10$) fields in three
  representative cold dark matter models.  Considering the current
  observational uncertainty of the frequency of such clustering at
  $z\sim3$, it would be premature to rule out the models, but the
  future spectroscopic surveys in a dozen fields could definitely
  challenge all the existing cosmological models {\it a posteriori}
  fitted to the $z=0$ universe.
\end{abstract}


\newpage

\section{Introduction}

While currently many cosmological models are known to be more or less
successful in reproducing the structure at redshift $z\sim0$, this may be
largely because there are still several degrees of freedom or {\it
cosmological parameters} to appropriately {\it fit} the observations
at $z\sim0$, including the density parameter, $\Omega_0$, the mass
fluctuation amplitude at the top-hat window radius of $8\himpc$,
$\sigma_8$, the Hubble constant in units of $100\kms {\rm Mpc}^{-1}$,
$h$, and even the cosmological constant $\lambda_0$. This kind of {\it
degeneracy} in cosmological parameters among viable models can be
broken by combining the data at higher $z$ (e.g., Jing \& Fang 1994;
Eke, Cole, \& Frenk 1996; Barbosa et al. 1996; Fan, Bahcall, \& Cen
1997; Kitayama, Sasaki \& Suto 1998). This is why recent very deep
surveys of galaxies, which have significantly advanced our
understanding of the properties and distribution of galaxies at high
redshifts, are equally important in constraining the underlying
cosmological models.

Recently Steidel et al. (1997; S97 hereafter) reported a discovery of
a highly significant concentration of galaxies on the basis of the
distribution of 78 spectroscopic redshifts in the range $2 \leq z \leq
3.4$ for photometrically selected ``Lyman break'' objects. This
discovery is quite important as discussed above, and therefore in this
{\it Letter} we focus on exploring its impact on models of cosmic
structure formation, specifically in the context of the cold dark
matter (CDM) cosmologies.

S97 have already considered implications of their discovery on the
basis of simple analytic models, and found that the objects are
associated with spatially highly biased dark halos of mass $M\sim
10^{12}M_\odot$ which confirms earlier suggestions based on
different grounds that the Lyman break galaxies are the progenitors of
present massive galaxies (Steidel et al. 1996; Mo \& Fukugita 1996;
Baugh et al. 1997).  We examine the theoretical impact of their
discovery in much greater detail using a large number of mock samples
from N-body simulations. Thus our analysis presented below properly
takes account of several important and realistic effects including (i)
the survey volume geometry, (ii) redshift-space distortion, (iii)
selection function of the objects, (iv) fully nonlinear evolution of
dark halos, and (v) finite sampling effect. As a result, we are able
to predict simultaneously the number density of such objects and their
clustering feature, and therefore can discuss the statistical
significance of their discovery.

\section{Simulation and selection procedure for ``Lyman break''
galaxies}

The simulations have been performed with the P$^3$M code (Jing \& Fang
1994) which now has been optimized for the vector processor.  We
consider three cosmological models summarized in Table 1, and run
three different realizations for each model. The initial conditions
are generated using the CDM transfer function of Bardeen et al. (1986)
which is fixed by the shape parameter $\Gamma=\Omega_0 h$. The three
models, the standard CDM (SCDM), a flat low-density CDM (LCDM), and an
open CDM (OCDM) we have chosen are consistent with the abundance of
low-redshift rich clusters (Kitayama \& Suto 1997), and are also very
close to those studied by S97.  All the simulations employ $256^3$
($\approx 17 $ million) particles in a (100\himpc)$^3$ comoving box
and start at redshift $z_i=36$. The gravitational softening length
$\epsilon$ is $39h^{-1}$kpc (comoving), and the mass of a simulation
particle $m_p$ is $1.7\times 10^{10} \Omega_0\himsun$, both of which
are small enough to resolve reliably the halos of Lyman break galaxies
$\sim 10^{12}\msun$.

We identify halos of galaxies using the Friend-Of-Friend (FOF) algorithm
with a bonding length $0.2$ times the mean particle separation, and
assume that each halo corresponds to one Lyman break galaxy. Although
this procedure is fairly idealized, it is definite in the sense that the
selected objects are characterized only by the threshold mass of the
halos, and provides a reasonable approximation in examining the
clustering feature of the resulting objects as a function of the halo
mass. The robustness of our result to the identification procedure
shall be discussed in \S 3. 

\begin{figure}
\begin{center}
  \leavevmode\psfig{figure=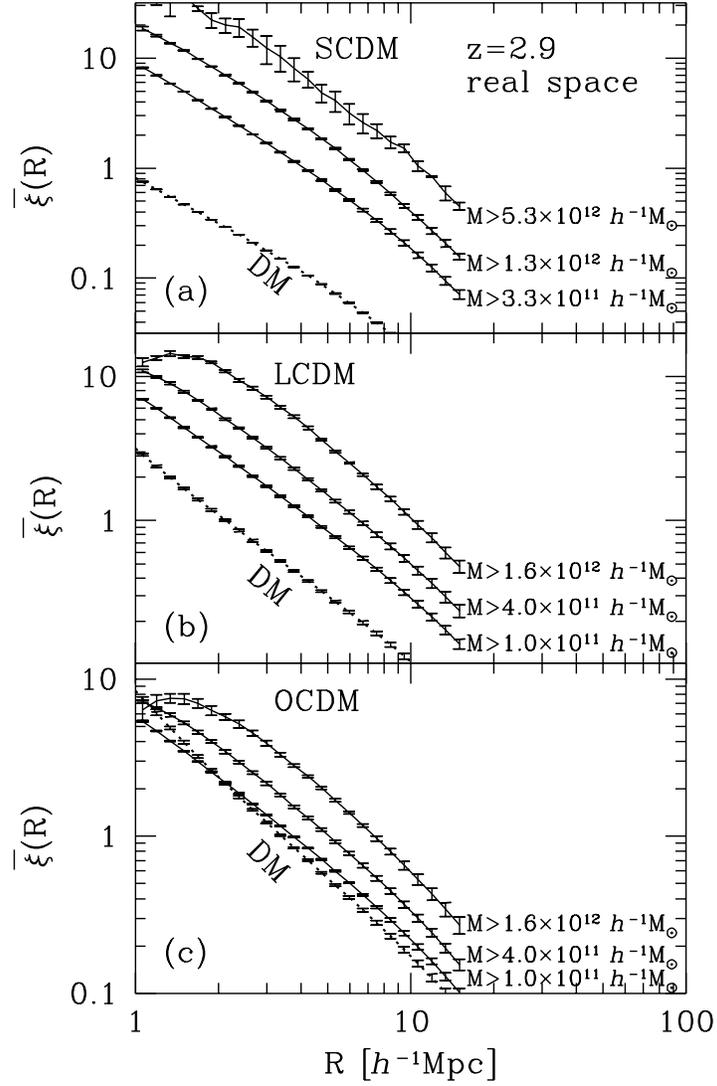,width=16cm}
\end{center}
\caption{Two-point correlation functions of halos at $z=2.9$ in real
space; different curves correspond to the different threshold masses
$M$ of the halos in (a) SCDM, (b) LCDM, and (c) OCDM models. Curves
labeled by DM correspond to the correlation functions of all
particles in the simulation.
\label{fig:xihalo}}
\end{figure}

Figure \ref{fig:xihalo} plots the volume-averaged two-point
correlation functions in real space $\bar{\xi} (R; M,z)$ of the halos
with mass larger than $M$ at redshift $z=2.9$. Here and throughout the
{\it Letter}, the error bars are calculated from the scatter among
three realizations for each model. Figure \ref{fig:xihalo} indicates
that over the scales of interest the correlation functions of halos
are enhanced approximately by a constant factor $b^2(M,z)$ relative to
those of the dark matter.  While this basic feature has been known
previously (e.g, Mo, Jing \& White 1996), Figure \ref{fig:xihalo}
clearly demonstrates the unprecedented quality (high mass resolution
and large volumes) of our simulated halo catalogues which is essential
in the statistical discussion below.  Figure \ref{fig:prob}a plots
this effective bias parameter for halos with mass larger than $M$,
$b(>M) \equiv \sqrt{ \bar{\xi} (R;M,z)/ {\bar \xi}_{\rm mass}(R;z)}$,
calculated at $R=7.5\himpc$ and $z=2.9$, where ${\bar \xi}_{\rm
mass}(R;z)$ is the volume-averaged correlation function of all {\it
particles} in the simulations.

\section{Statistical analysis with mock samples}

The field, SSA22, reported by S97 covers $8.74\times 17.64$ arcmin$^2$
sky area at $\alpha = 22^h17^m$ and $\delta=+00^\circ15'$. There are a
total of 181 objects with magnitude ${\cal R} \le 25.5$ which satisfy
their color selection criteria. From their follow-up spectroscopic
observations, it turns out that about 70 percent of the objects
(i.e. about 130 objects in the field) are galaxies at $2\le z\le
4$. They have measured redshift for 67 galaxies, which shows a strong
concentration of 15 galaxies in one $\Delta z = 0.04$ bin at
$z=3.1$. In this section, we perform the statistical analysis with
mock samples to examine the extent to which the CDM models can
reproduce such a concentration. In generating the mock samples, we
adopt a redshift selection function which is spline-interpolated
from the redshift distribution histogram of Figure 2 of Pettini et al.
(1997) and then normalized so as to have 130 galaxies in SSA22. Our
resulting redshift selection function is approximately the same as
that given by S97 when normalized to the same number of galaxies.

Between redshifts 2 and 4, each of our realizations has 7 outputs with
time intervals $\Delta \ln(1+z)=0.1$. For a given threshold of the halo
mass, we combine these outputs at different epochs with periodic
replications to generate the mock samples of the SSA22 field.  The
peculiar velocity and cosmological distortion effects on the distance -
redshift relation (Matsubara \& Suto 1996) are taken into account
self-consistently in computing each redshift of the mock galaxies.  The
sky area of each mock field is fixed to be the same as that of SSA22,
and halos inside each mock field are selected randomly according to
the redshift selection function described above.  Finally we randomly
pick up 67 objects to mimic their spectroscopic observation. A total
of 12,000 mock samples are generated for a given threshold of the halo
mass in each model.

\begin{figure}
\begin{center}
  \leavevmode\psfig{figure=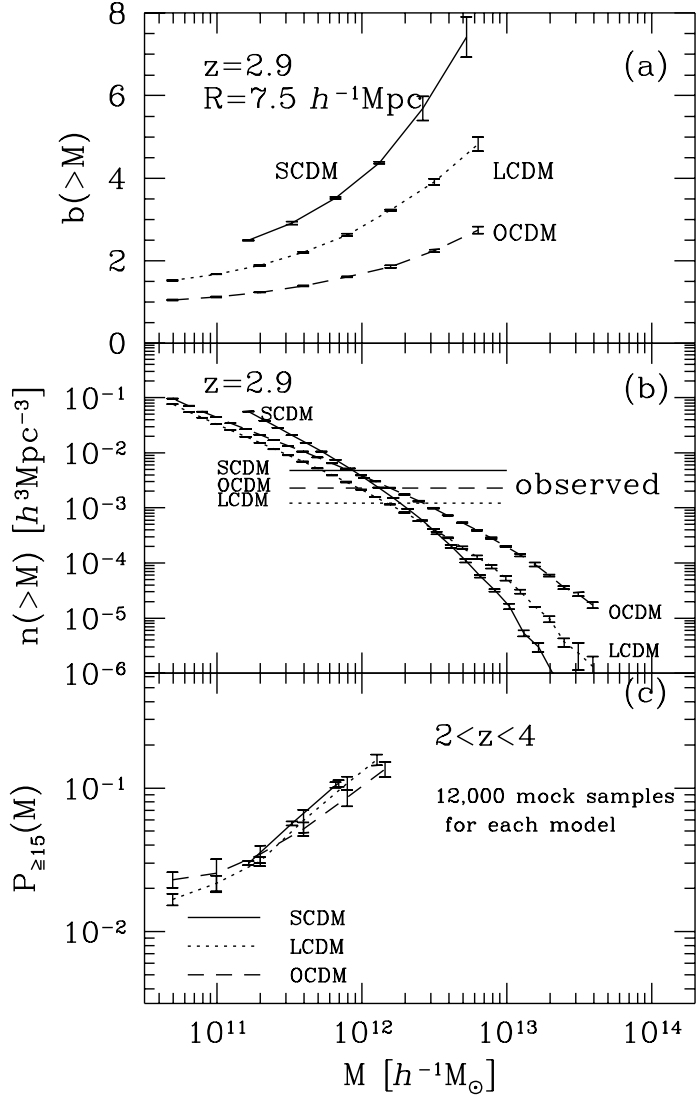,width=16cm}
\end{center}
\caption{Statistics of the halos as a function of the threshold mass;
(a) bias parameter of halos at $z=2.9$; (b) mean number density of
halos in the simulations compared with the observed one of Lyman break
galaxies; (c) probability of finding a concentration of at least 15
halos in a bin of $\Delta z=0.04$.
\label{fig:prob}}
\end{figure}

The mean number density of halos of mass larger than $M$ (without
applying the selection function) is plotted in Figure \ref{fig:prob}b.
Three horizontal lines indicate the observed number density of Lyman
break galaxies corresponding to our three model parameters. Note that
the observed number density should be regarded as a strictly lower
limit since some fraction of the galaxies might have been unobserved
due to the selection criteria.  Since the density of halos falls below
the observed one for $M>M_{max}$, we vary the threshold mass of the
halos from $10 m_p$ up to $M_{max}$ in considering the statistical
significance of the clustering.  We use a simple algorithm to identify
a galaxy concentration in the mock sample following the procedure of
S97; for each mock sample, we count galaxies within redshift bins of
$\Delta z=0.04$ centered at each galaxy and identify the redshift bin
with the maximum count as the density concentration.  Then we compute
the probability $P_{\gequal 15}(M)$ that a mock sample has a
concentration with at least 15 galaxies. Figure \ref{fig:prob}c
indicates that the probability $P_{\gequal 15}(M)$ increases with the
halo mass as expected, since the clustering is stronger for more
massive halos.  For the halo mass $M\simlt 10^{11}\himsun$,
$P_{\gequal 15}(M)$ is less than 5 percent for all the three
models. The probability $P_{\gequal 15}(M)$ can increase to 15\% for
LCDM and OCDM, and to 10\% for SCDM at $M_{max}$.

\begin{figure}
\begin{center}
  \leavevmode\psfig{figure=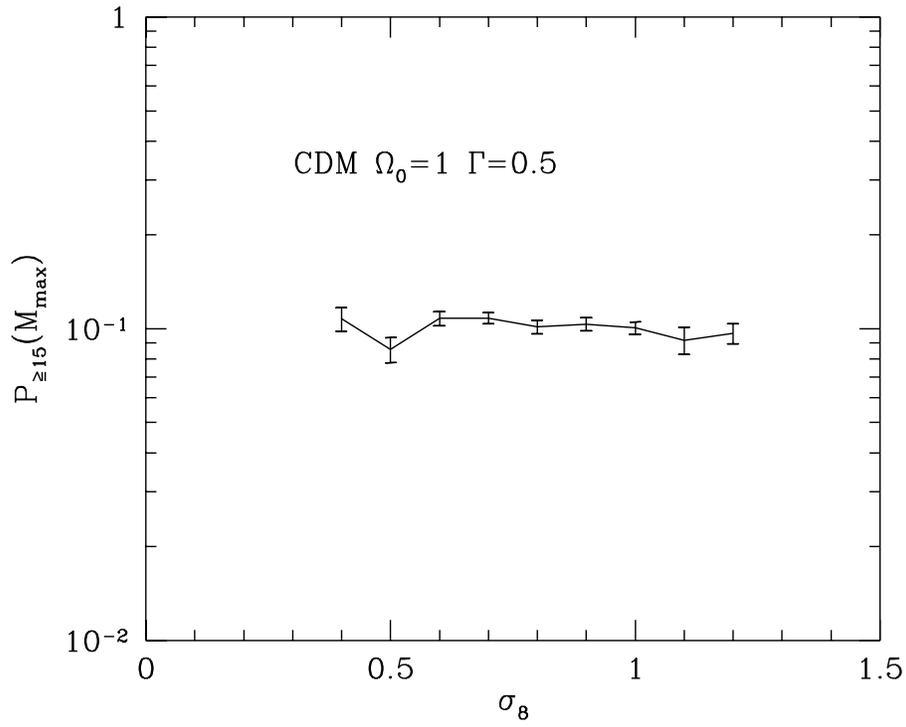,width=16cm}
\end{center}
\caption{ Maximum probability  $P_{\protect\gequal\protect 15}(M_{max})$
in SCDM ($\Gamma=0.5$) as a function of $\sigma_8$.
\label{fig:psigma8}}
\end{figure}

The above conclusion might be dependent on the value of $\sigma_8$
which we fixed for each model.  In the case of SCDM with $\Gamma=0.5$,
we can study its dependence by analyzing the simulation outputs at
different epochs. Figure \ref{fig:psigma8} plots the probability
$P_{\gequal 15}(M_{max})$, i.e., the maximum probability as a function
of $\sigma_8$. The probability is about 10\% for $\sigma_8\le 1.2$
which is the value suggested by the 4yr COBE data for the Harrison -
Zel'dovich primordial spectrum in this model.

Since the identification of ``galaxies'' from pure N-body simulations
is not totally unambiguous, we have carefully examined if our results
are sensitive to the particular identification procedure adopted here.
First we tried another identification scheme in which those spherical
regions around each potential minimum with overdensity above
$178\Omega_0^{-0.6}$ are defined as halos.  It has been shown that
this scheme removes some problems of the FOF method (e.g., Jing \&
Fang 1994). With such selected dark halos, we repeated our procedure
and found that the resulting $P_{\gequal 15}(M)$ agrees within 2\%
with that based on the FOF halos.  Thus, provided that one dark halo
corresponds to one galaxy, our results presented above are quite
robust.

Another critical question is the over-merging effect: one virialized
halo in principle may harbor more than one galaxy. If this were the
case with the observed Lyman-break galaxies, our analysis would
under-predict the probability $P_{\gequal 15}$. However this seems to
be very unlikely; among the 15 galaxies in the concentration of S97,
there are only two pairs of galaxies with a sky separation less than
$1'$ ($\approx 1\himpc$ in comoving). Both pairs have a radial
velocity difference about $2000 \kms$ (in the rest-frame at $z=3.1$),
so it is not likely that each pair is in the same virialized halo
(note that at $z$ about 3, rich clusters with a velocity dispersion
about $2000\kms$ are very rare in CDM models). We have also checked
our mock samples directly and found that each high concentration on
average has 3 pairs of mock galaxies closer than $1'$ in all the three
models we considered, which agrees with the observation. This
indicates that if we further break virialized halos into mock
galaxies, we would have too many close pairs. To be more specific, we
have experimented this procedure for the OCDM simulations by
identifying halos with FOF at a much earlier epoch $z\approx 6$,
placing mock galaxies to the center particle of each halo, and
assuming all these mock galaxies survive until the observed redshift
(i.e., no later merger). In this case the probability $P_{\gequal
15}(M_{max})$ is nearly doubled, but there are on average about 7
pairs of galaxies closer than $1'$ in each high concentration, which
is much higher than that observed. Therefore it appears that the
over-merging effect does not significantly bias our result.

Considering the fact that S98 observed the concentration in their
{\it first} densely-sampled field, it seems that the CDM models are
only marginally consistent with the observation depending on its
statistical significance.

\section{Conclusions}

We have carried out a detailed analysis of the statistical
significance of a concentration of galaxies at $z \sim 3$ discovered
by S97 based on mock samples constructed from a series of
N-body simulations. The observed number density of Lyman break
galaxies at $z\sim3$ implies that they correspond to systems with dark
matter halos of $\simlt 10^{12}M_\odot$, as suggested by Steidel et
al. (1996).  While the clustering of such objects is naturally biased
with respect to dark matter, the predicted bias $1.5\sim 3$ is not
large enough to be reconciled with such a strong concentration of
galaxies at $z \sim 3$ if one similar structure is found per one field
on average; we predict one similar concentration approximately per ten
fields in SCDM and per six fields in LCDM and OCDM.  Our procedure is
definite and reliable in the sense that it is derived from the mock
samples which adopt all the realistic effects relevant to the
statistical discussion.  Our conclusions are qualitatively consistent
with those of S97, but in fact much more stringent mainly because
we simultaneously consider the constraints from both the observed
number density of objects and their clustering; the predicted number
density of the more biased halos, which is consistent with the
strong concentration, is too small to be compatible with the observed
one.  Although it would be premature to rule out the models
discussed here, we have clearly demonstrated the importance of deep
surveys of galaxies in constraining the cosmological models;
future spectroscopic surveys in a dozen fields (Pettini et al. 1997)
could challenge all the existing cosmological models {\it a
posteriori} fitted to the $z=0$ universe.

\bigskip 
\bigskip 

We would like to thank Gerhard B\"orner for a carefully reading of the
manuscript and useful comments, and an anonymous referee for an
important comment which leads to a discussion at the end of section 3.
Numerical computations were carried out on VPP300/16R and VX/4R at the
Astronomical Data Analysis Center of the National Astronomical
Observatory, Japan, as well as at RESCEU (Research Center for the
Early Universe, University of Tokyo) and KEK (National Laboratory for
High Energy Physics, Japan). Y.P.J. gratefully acknowledges the
postdoctoral fellowship from Japan Society for the Promotion of
Science. This research was supported by the Grants-in-Aid of the
Ministry of Education, Science, Sports and Culture of Japan
No.07CE2002 to RESCEU, and No.96183 to Y.P.J., and by the
Supercomputer Project (No.97-22) of High Energy Accelerator Research
Organization (KEK).

\newpage
\begin{table}[h]
\begin{center}
  Table~1.\hspace{4pt} Simulation model parameters \\ 
\end{center}
\vspace{6pt}
\begin{center}
\begin{tabular}{ccccccc}
\hline\hline\\[-6pt]
Model & $\Omega_0$ &  $\lambda_0$  
&  $\Gamma$ &   $\sigma_8$ & $m_p$ ($\himsun$) \\ 
[4pt]\hline \\[-6pt]
SCDM & 1.0  & 0.0 & 0.5 & 0.6 & $1.7\times 10^{10}$\\
OCDM & 0.3  & 0.0 & 0.25 & 1.0 &$5.0\times 10^{9}$\\
LCDM & 0.3  & 0.7 & 0.21 & 1.0 &$5.0\times 10^{9}$\\
\hline
\end{tabular}
\end{center}
\end{table}

\newpage

\parskip2pt
\newpage
\centerline{\bf REFERENCES}
\bigskip

\def\pp{\par\parshape 2 0truecm 15.5truecm 1truecm 14.5truecm\noindent}
\def\apjpap#1;#2;#3;#4; {\pp#1, {#2}, {#3}, #4}
\def\apjbook#1;#2;#3;#4; {\pp#1, {#2} (#3: #4)}
\def\apjppt#1;#2; {\pp#1, #2.}
\def\apjproc#1;#2;#3;#4;#5;#6; {\pp#1, {#2} #3, (#4: #5), #6}

\apjpap Barbosa, D., Bartlett, J.G., Blanchard, A., \& Oukbir,
J. 1996;A\& A;314;13;
\apjpap Bardeen, J. M., Bond, J. R., Kaiser, N. \& Szalay, A. S. 1986;
  ApJ;304;15;
\apjppt Baugh, C.M., Cole, S., Frenk, C.S., \&  Lacey,
C.G. 1997;MNRAS, submitted (astro-ph/9703111);
\apjpap Eke, V. R., Cole, S., \& Frenk, C. S. 1996;MNRAS;282;263;
\apjpap Fan,X., Bahcall, N.A. \&  Cen, R. 1997;ApJ;490;L123;
\apjpap Jing, Y.P., \& Fang, L.Z. 1994;ApJ;432;438;
\apjppt Kitayama, T., Sasaki, S. \& Suto,
Y. 1998;Pub.Astron.Soc.Japan, in press (astro-ph/9708088);
\apjpap Kitayama, T., \& Suto, Y. 1997;ApJ;490;557;
\apjpap Mo,H.J., \& Fukugita, M. 1996;ApJ;467;L9;
\apjpap Mo,H.J., Jing, Y.P., \& White, S.D.M. 1997;MNRAS;284;189;
\apjpap Matsubara, T., \& Suto, Y. 1996;ApJ;470;L1;
\apjppt Pettini, M., Steidel, C.C., Adelberger, Kellogg, M., K.L.,
Dickinson, M., \& Giavalisco, M. 1997; to appear in the proceedings of
`ORIGINS', ed. J.M.  Shull, C.E. Woodward, and H. Thronson, (ASP
Conference Series) (astro-ph/9708117);
\apjpap Steidel, C.C.,Giavalisco,
M., Pettini, M., Dickinson, M., \& Adelberger, K.L. 1996;ApJ;
462;L17;
\apjppt Steidel, C.C., Adelberger, K.L., Dickinson, M., Giavalisco,
M., Pettini, M., \& Kellogg, M. 1997;ApJ, submitted (astro-ph/9708125:
S97);
\newpage

\end{document}